\documentclass[conference]{IEEEtran}
\IEEEoverridecommandlockouts
\usepackage{cite}
\usepackage{amsmath,amssymb,amsfonts}
\usepackage{algorithmic}
\usepackage{graphicx}
\usepackage{textcomp}
\usepackage{xcolor}
\usepackage{stfloats}
\usepackage{float} %设置图片浮动位置的宏包
\usepackage{subfigure} %插入多图时用子图显示的宏包
 \usepackage{fancyhdr} % 添加页眉页脚
\usepackage{CJKutf8}
\usepackage[ruled,linesnumbered]{algorithm2e}
\usepackage{amsmath} 
\usepackage [utf8] { inputenc } 
\usepackage{booktabs}
\usepackage { graphics } \usepackage { placeins }
\usepackage[misc]{ifsym}
\def\BibTeX{{\rm B\kern-.05em{\sc i\kern-.025em b}\kern-.08em
    T\kern-.1667em\lower.7ex\hbox{E}\kern-.125emX}}
\begin{document}

\title{AMD-DBSCAN: An Adaptive Multi-density DBSCAN for datasets of extremely variable density}

\author{
\IEEEauthorblockN{
Ziqing Wang\textsuperscript{1}, Zhirong Ye\textsuperscript{1}, Yuyang Du\textsuperscript{1}, Yi Mao\textsuperscript{1},  Yanying Liu\textsuperscript{1}, Ziling Wu\textsuperscript{2}, Jun Wang\textsuperscript{3}
\thanks{\rule[0pt]{8.55cm}{0.05em}}
\thanks{This work was supported by Key-Area R\&D Program of Guangdong (2019B010135002), Innovative \& Enterprising Team of Zhuhai (2019ZHCDGY07), and University-enterprise project (K20-76220-002). }
\thanks{ \textsuperscript{3} Contact Author }
}
\IEEEauthorblockA{
School of Microelectronics Science and Technology \\
\textsuperscript{1}Sun Yat-sen University\\
\{wangzq37, yezhr6, duyy27, maoy28, liuyy336\}@mail2.sysu.edu.cn\\
\textsuperscript{2}University of Nottingham\\
psxzw11@nottingham.ac.uk\\
\textsuperscript{3}Zhuhai Jieli Technology Co.,Ltd\\
jimmy.junwang@gmail.com
}
}

\maketitle

\begin{abstract}
DBSCAN has been widely used in density-based clustering algorithms. However, with the increasing demand for Multi-density clustering, previous traditional DSBCAN can not have good clustering results on Multi-density datasets. In order to address this problem, an adaptive Multi-density DBSCAN algorithm (AMD-DBSCAN) is proposed in this paper. An improved parameter adaptation method is proposed in AMD-DBSCAN to search for multiple parameter pairs (i.e., Eps and MinPts), which are the key parameters to determine the clustering results and performance, therefore allowing the model to be applied to Multi-density datasets. Moreover, only one hyperparameter is required for AMD-DBSCAN to avoid the complicated repetitive initialization operations. Furthermore, the variance of the number of neighbors (VNN) is proposed to measure the difference in density between each cluster. The experimental results show that our AMD-DBSCAN reduces execution time by an average of 75\% due to lower algorithm complexity compared with the traditional adaptive algorithm. In addition, AMD-DBSCAN improves accuracy by 24.7\% on average over the state-of-the-art design on Multi-density datasets of extremely variable density, while having no performance loss in Single-density scenarios. Our code and datasets are available at \url{https://github.com/AlexandreWANG915/AMD-DBSCAN}.
\end{abstract}

%\begin{IEEEkeywords}
%DBSCAN, Parameter Adaptation, Multi-denstiy
%\end{IEEEkeywords}

\section{Introduction}
DBSCAN\cite{esterDensitybasedAlgorithmDiscovering1996a} is one of the most widely used density-based clustering methods in data mining\cite{fayyadDataMiningKnowledge1996}. Since objects within a cluster are similar and objects in different clusters are dissimilar \cite{hanDataMiningConcepts2001}, DBSCAN is able to classify the objects with the similar characteristics into one cluster\cite{cooperComparativeSurveyVANET2016} and to identify clusters of different shapes and sizes accurately from noise\cite{aryaOptimizedApproachDensity2014}. Due to its specialties, it is widely used in various fields such as ship detection\cite{langShipDetectionHighresolution2019}, wafer classification\cite{jinNovelDBSCANbasedDefect2019}, astronomy\cite{sanderDensitybasedClusteringSpatial1998}, robotics\cite{bewleyRealtimeVolumeEstimation2011}and disease diagnosis\cite{pasinUsageKernelMeans2015}. 

However, traditional DBSCAN has two drawbacks. Firstly, DBSCAN clustering relies on two parameters (i.e., Eps and MinPts). For a point, $Eps$ is the radius of the circle whose center is that point. All the points in that circle could be considered as neighbors of that point. $MinPts$ is a threshold value that can be utilized to search core points during the process of DBSCAN clustering.  However, in most cases, the dataset has a dimensionality greater than three dimensions, which leads to the inability of the visualization. Therefore, it is hard to determine these two parameters, and it would take a lot of time and effort to modify the parameters artificially. And when the dimensionality of the data is high, there will be a curse of dimensionality\cite{keoghCurseDimensionality2017}, leading to a lower clustering accuracy. 

Furthermore, traditional DBSCAN only offers a single parameter pair and it has a low accuracy on Multi-density datasets of extremely variable density. To be more specific, the density of each cluster varies greatly. Giving parameters at the density of the sparse clusters would lead to merging similar datasets, while giving parameters at the density of the dense clusters would result in many clusters with less density being incorrectly identified as noise. Therefore, traditional DBSCAN that uses only a fixed parameter pair could not give good clustering results on Multi-density datasets where different regions have various densities.

Some algorithms (e.g., YADING\cite{dingYadingFastClustering2015}) propose Multi-density DBSCAN to address this problem. Specifically, the data points with the highest density are firstly clustered, then the data points that have been clustered are saved, and then the remaining points that have not been clustered are continued to be clustered. Repeat the above process until all points are clustered. This means that different density clusters should utilize different parameter pairs. To obtain multiple parameter pairs, most of these algorithms utilize the $k_{dis}$ curve. To determine a $k_{dis}$ curve, the Euclidean distance matrix, a square matrix containing the euclidean distances between the elements of a set, is supposed to be obtained first and sorted in ascending order. The $k_{dis}$ curve is comprised of the values of the $k^{th}$ column of this sorted distance matrix. And $k$ means $k^{th}$ neighbor of a point and different $k$ values determine different curves, resulting in different parameter pairs. However, these methods mostly use a fixed $k$, which is not well based on the distribution properties of the dataset. In addition, to search for multiple candidate parameter pairs, they introduce additional hyperparameters that are required to be set manually.

To enable DBSCAN to be applied to Multi-density datasets of extremely variable density, AMD-DBSCAN is proposed. To the best of our knowledge, the proposed approach in this paper is a novel exploration of the $k_{dis}$ value, which is the $k^{th}$ nearest neighbor of a point, to search for multiple parameter pairs matching the distribution of the dataset.

The contributions of this paper are summarized as follows:
\begin{enumerate}
\item  An improved parameter adaptation method is proposed to locate the adaptive $k$ based on the distribution of the datasets and the binary search algorithm is utilized to speed up this adaptive process.
\item  A new method for utilizing the $k_{dis}$ value is proposed to search for multiple candidate $Eps$.
\item  An adaptive Multi-density clustering algorithm is proposed to provide matching parameter pairs (i.e., $Eps$ and $MinPts$) for each density cluster.
\end{enumerate}

\section{RELATED WORK}
The clustering results of DBSCAN are significantly affected by two parameters (i.e., $Eps$ and $MinPts$) and the distribution of datasets. As for parameters, larger $MinPts$ and smaller $Eps$ lead to incomplete clustering (many core points are considered as noise). In contrast, small $MinPts$ and large $Eps$ result in over-clustering (two or more clusters of points are clustered into a single cluster). In the case of datasets' distribution, only one pair of $Eps$ and $MinPts$ is unlikely to cope with the Multi-density datasets whose points are not evenly distributed. In this section, previous work on these two aspects is discussed and compared with AMD-DBSCAN.

\subsection{Configuration of parameters}

To obtain appropriate parameters for DBSCAN, many approaches have been proposed. These methods can be divided into two categories, adapting one of the two parameters (i.e., $Eps$ and $MinPts$) and configuring both of them.

\subsubsection{\textbf{Single Parameter}}
 Reversing the nearest neighbor has been proposed by method \cite{bryantRNNDBSCANDensitybasedClustering2017} to estimate the density around a point for adapting $MinPts$ automatically. Besides, the $k_{dis}$ curve, which is comprised of the distance of points of the dataset and their k-nearest neighbors in ascending order, is leveraged by \cite{mistryAEDBSCANAdaptiveEpsilon2021, dingYadingFastClustering2015, liRobustRapidClustering2018} to adapt appropriate $Eps$ by locating the flat part of the curve. However, the analysis of the $k_{dis}$ curve proposed by \cite{dingYadingFastClustering2015, liRobustRapidClustering2018} requires another two parameters and is with a relatively large complexity $(O(nlog(n)))$. ISB-DBSCAN\cite{lvEfficientScalableDensitybased2016} algorithm utilizes an input parameter k as the number of the nearest neighbor to reduce the input parameters of DBSCAN. AA-DBSCAN\cite{kimAADBSCANApproximateAdaptive2019a} algorithm utilizes the approximate adaptive $Eps$ for each density. Hence, it can find the clusters in the Multi-density datasets.

However, all the approaches above can only adapt one parameter automatically. The advantage of the AMD-DBSCAN over the above algorithms is that our method can adapt both parameters at the same time.

\subsubsection{\textbf{Multiple Parameters}}
The following method is proposed to provide two parameters. In GMDBSCAN \cite{xiaoyunGMDBSCANMultidensityDBSCAN2008}, centers of clusters are generated by GD to adapt both two parameters. However, it has a lower accuracy compared to AMD-DBSCAN when dealing with the Multi-density datasets.

\subsection{Properties of Datasets}
\subsubsection{\textbf{Multi-density Datasets}}
Many algorithms are proposed for tackling the situation of Multi-density datasets with unevenly distributed points. For instance, VDBSCAN\cite{liuVDBSCANVariedDensity2007} proposed to automatically customize clustering parameters for different density regions. Furthermore, AEDBSCAN\cite{mistryAEDBSCANAdaptiveEpsilon2021} improves the $k_{dis}$ curve by using the second-order difference method to improve the accuracy of the Multi-density clustering. Also, YADING\cite{dingYadingFastClustering2015} estimates the density by determining the $Eps$ based on VDBSCAN\cite{liuVDBSCANVariedDensity2007}, and it optimizes the clustering speed. DVBSCAN \cite{ramDensityBasedAlgorithm2010} can deal with local density variation within a cluster, but it cannot determine parameters automatically. HDBSCAN\cite{campelloDensitybasedClusteringBased2013} is a hierarchical clustering method that allows it to perform well on Multi-density datasets. However, it has a long execution time.

Since AMD-DBSCAN can adapt a parameter pair for each layer during the Multi-density DBSCAN process, it can achieve good clustering results on Multi-density datasets.

\subsubsection{\textbf{Large-scale Datasets}}
Some algorithms are proposed to be applied under large-scale datasets. Some methods \cite{heMrdbscanEfficientParallel2011a, yuCludoopEfficientDistributed2015, mohammedApplicationsMapReduceProgramming2014} use a partitioning strategy and a distributed structure that allow them to be applied to large-scale datasets. SDBSCAN algorithm \cite{zhouCombiningSamplingTechnique2000} combines sampling techniques with DBSCAN for clustering large spatial databases. In IDBSCAN\cite{borahImprovedSamplingbasedDBSCAN2004a} method, the greater I/O cost and memory requirements involved in clustering are addressed by using marked boundary objects to directly scale the computation without the need for actual dataset selection. 

However, all of these methods above have a low accuracy on Multi-density datasets. AMD-DBSCAN can have good performance in this scenario because it can adapt two parameters according to the distribution of large-scale datasets.

\section{PROPOSED ALGORITHM}
\subsection{Overview}
\begin{figure*}[htbp]
  \centering
  \includegraphics[width=\linewidth]{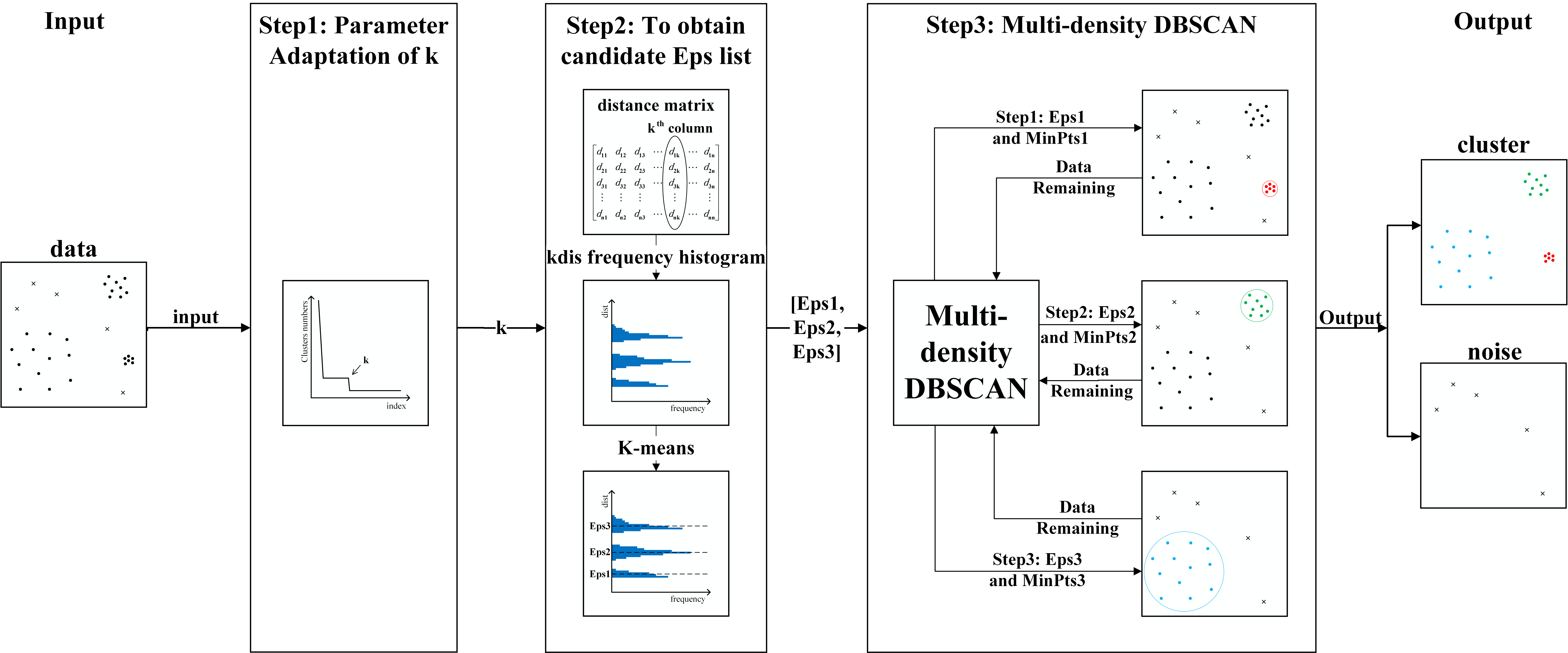}
  \caption{Framework of AMD-DBSCAN}
  \label{total}
\end{figure*}

In this section, the details of the three steps of AMD-DBSCAN are analyzed and the complexity of the overall algorithm is given at the end. Figure \ref{total} gives the information of the framework proposed in this paper. 

\begin{enumerate}
\item \textbf{Parameter Adaptation of k:} To achieve Multi-density clustering on the datasets of extremely variable density, it is first necessary to obtain $k$ required to determine the $k_{dis}$ value. An improved parameter adaptation method is proposed to locate $k$. The spatial distribution properties of the dataset itself are utilized to generate a list of candidate $Eps$ and $MinPts$ parameters, which are required by DBSCAN. The $Eps$ and $MinPts$ parameter lists are sequentially input into the DBSCAN to obtain the number of clusters. After that, by using the binary search algorithm, $Eps$ and $MinPts$ pairs that are the most match the distribution of the dataset for clustering is selected, where $MinPts$ is used as $k$ required for the next step.

\item \textbf{To obtain Candidate Eps List:} The $k_{dis}$ value, which is defined as $k^{th}$ neighbor of a point, is determined by $k$ derived from the last step. And the $k_{dis}$ frequency histogram can be obtained by a series of $k_{dis}$ values arranged in ascending order. The K-means algorithm is utilized to cluster the $k_{dis}$ values to obtain $N$ candidate $Eps$, where $N$ refers to the number of peaks by directly observing the $k_{dis}$ frequency histogram.

\item \textbf{Multi-density Clustering:} Multi-density DBSCAN can perform well on Multi-density datasets and it is introduced in this study. Each $Eps$ in the obtained $CandidateEpsList$ is sorted in ascending order and used to generate the corresponding $MinPts$. The parameters of each layer are input into Multi-density clustering respectively to obtain the final results.
\end{enumerate}

\subsection{Parameter Adaptation of k}
There are many algorithms \cite{dingYadingFastClustering2015, liRobustRapidClustering2018} that utilize $k_{dis}$ value, which represents the $k^{th}$ nearest neighbor of each point, to obtain the important parameter $Eps$ for clustering. They require manual input of the parameter $k$. And our model proposes a novel exploration of $k_{dis}$ value using the $k_{dis}$ frequency histogram. The clustering results depend directly on the selection of the parameter $k$. 

According to our experiments shown in the latter section, a small $k$ leads to a large number of clusters (what should be one large cluster is clustered into many small clusters), while a large $k$ results in a small number of clusters (what should be multiple clusters are clustered into one cluster). Therefore, $k$ affects the clustering results significantly.

To avoid the randomness of manually setting $k$ during clustering, a parameter adaptation algorithm is proposed in this study to search for $k$. This approach adapts a $k$ that responds to the distribution properties of the dataset. Our subsequent experiments demonstrate that this adaptive $k$ is an important guide for determining the $k_{dis}$ value, making parameter pairs can have good performance for Multi-density clustering. The specific process of parameter adaptation of $k$ is as follows.

\subsubsection{\textbf{To compute Eps List}}
First of all, the Euclidean distance matrix  $\text{DIST}_{n \times n}$ of the dataset $D$ is calculated:
\begin{equation}
\text{DIST}_{n \times n}=\{dist(i, j),  1 \leq i \leq n, 1 \leq j \leq n\}
\end{equation}
where $n=|D|$ represents the number of points in the dataset $D$, $\text{DIST}_{n \times n}$ is a symmetric matrix with $n$ rows and $n$ columns, and each element represents the Euclidean distance from $i^{th}$ point to $j^{th}$ point in the dataset $D$.

By arranging each row in $\text{DIST}_{n \times n}$ in ascending order, a $\text{SORTED\_DIST}_{n \times n}$ is obtained. The distance data of the $k^{th}$ column in $\text{SORTED\_DIST}_{n \times n}$ is noted as vector $D_{k}$. And $\overline{D_{k}}$ is obtained by averaging the data in the vector $D_{k}$, which yields a list of $Eps$ as shown in Equation \ref{eps}.

\begin{equation}
EpsList=\left\{\overline{D_{k}}, 1 \leq k \leq n\right\}
\label{eps}
\end{equation}

The whole process is encapsulated as a function $obtainEpsList(data)$ which will be used in the pseudo-code. The input is the dataset $D$ and the output is $EpsList$.

\subsubsection{\textbf{To compute MinPts List}}
For a given $EpsList$, each $Eps$ value in it is utilized to calculate a $MinPts$, which is the number of neighbors corresponding to each $Eps$. And $MinPtsList$ consists of multiple $MinPts$, as shown in Equation \ref{minpts1} and Equation \ref{minpts2}.
\begin{equation}
MinPtsList=\left\{MinPts_{j}, 1 \leq j \leq n\right\}
\label{minpts1}
\end{equation}

\begin{equation}
MinPts_{j}=\frac{1}{n} \sum_{i=1}^{n} Point_{i}
\label{minpts2}
\end{equation}

where $Point_i$ is the number of neighbors of the $i^{th}$ point within the range of $Eps_{j}$, and $n$ represents the number of points in the dataset $D$.

The whole procedure above is encapsulated as a function $obtainMinPtsList(data, Epslist)$. The input are the dataset $D$ and $EpsList$ and the output is $MinPtsList$.

\begin{figure}[htbp]

\centering
\subfigure[cluster number]{
\label{num}
\includegraphics[width=3.9cm]{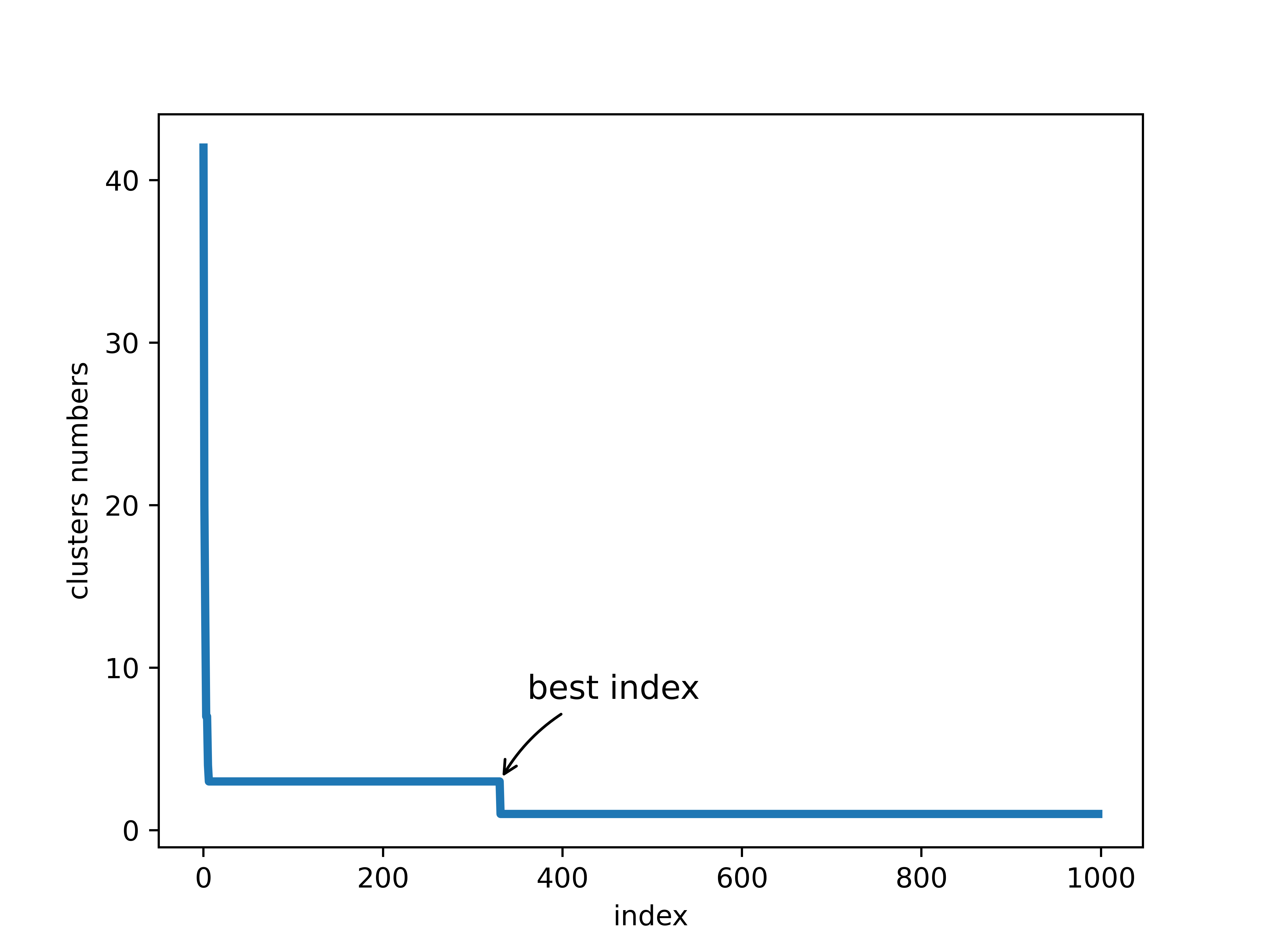}
%\caption{fig1}
}
\quad
\subfigure[NMI score]{
\label{score}
\includegraphics[width=3.9cm]{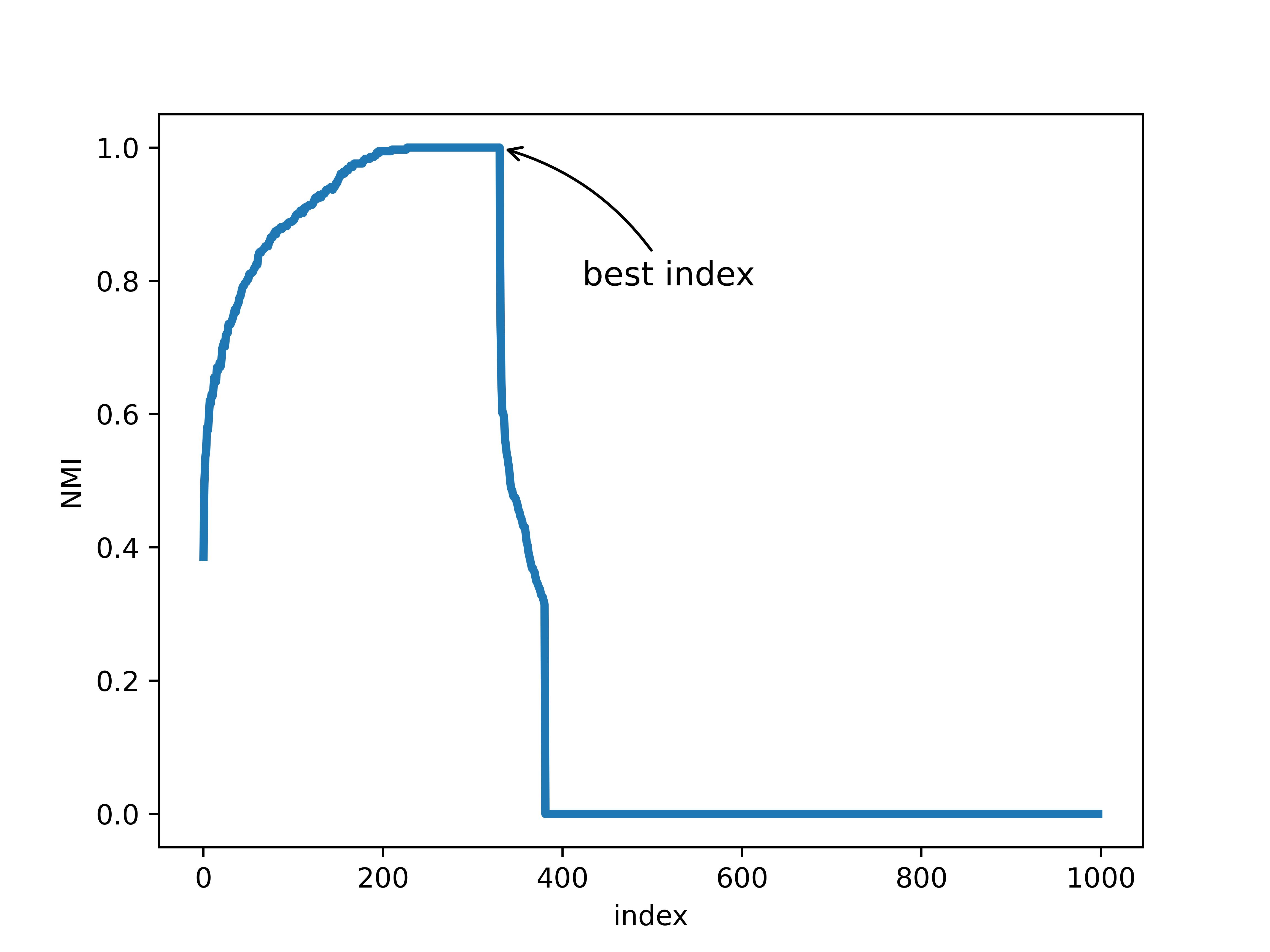}
}
\caption{\textbf{Variation of the clusters numbers and NMI score.} (a)Best index is the maximum index when the number of clusters is stable for the first time. (b)The index of the highest NMI score is exactly equal to the best index.}
\end{figure}

\subsubsection{\textbf{To locate the adaptive k}}
The algorithm is based on two conclusions obtained by analyzing the image of the number of clusters and the image of the effect of clustering.
\begin{enumerate}
\item $EpsList$ is in ascending order, which means that as the $index$ increases, $Eps$ increases. An experiment is designed in which each pair of $Eps$ and $MinPts$ parameters are input sequentially into DBSCAN to obtain the number of clusters. By observing the change in the number of clusters with index, our experimental results show that the number of clusters decreases monotonically with increasing $index$, as shown in Figure \ref {num}. That is because as $Eps$ increases, more points in the dataset are clustered into one class, and thus the number of clusters is monotonically decreasing.
\item When the number of clusters is stable for the first time, which means the number of clusters is the same three times in a row shown in the Figure \ref{num}, the point with the largest index in this stable part of the curve corresponds to the best clustering. Normalized Mutual Information (NMI)\cite{danonComparingCommunityStructure2005} is utilized to judge the clustering effect. An experiment is designed whereby the NMI of each DBSCAN result is recorded. By analyzing the change in NMI with index, our experimental results shown in Figure \ref{score} indicate that the NMI increases until reaches the largest index when the number of clusters is stable, which is noted as the best index.
\end{enumerate}

Based on the above conclusions, an improved parameter adaptation method is proposed. The specific implementation is shown below. The $EpsList$ and $MinPtsList$ are chosen as the $Eps$ candidates and $MinPts$ parameters, respectively. And then each parameter pair is input into DBSCAN for clustering to obtain the number of clusters. As illustrated in Algorithm \ref{algorithm1}, when the number of clusters is the same three times in a row for the first time, the clustering result is considered to be stable. The number of clusters $n$ is considered to represent the number of true clusters of the dataset. And the best index where the NMI value is the greatest is the largest index when the number of clusters is equal to $n$.

According to the first conclusion, the number of clusters decreases monotonically with $index$. To avoid traversing all combinations of parameters and reduce the time significantly, the binary search algorithm is utilized to locate the best index. The search array is from the first index equal to $n$ to the last index. The search process starts from the middle index. If the number of clusters corresponding to the middle index is exactly $n$, the search starts from the right half of the array. If it is less than $n$, the search starts from the left half of the array. Repeat this process until the largest index that the number of clusters is equal to $n$ is located, that is, the best index. Finally, the $MinPts$ corresponding to the best index in the $MinPtsList$ is our adaptive $k$.

The pseudo-code of locating $k$ is shown in Algorithm \ref{algorithm1}.

 \SetAlFnt{\small}
\begin{algorithm}[ht]
\SetKwData{Left}{left}\SetKwData{This}{this}\SetKwData{Up}{up}
\SetKwFunction{Union}{Union}\SetKwFunction{FindCompress}{FindCompress}
\SetKwInOut{Input}{input}\SetKwInOut{Output}{output}
\caption{Parameter Adaptation of k}\label{algorithm1}
\Input {$data$}
\Output{$k$}
$Eps \leftarrow obtainEpsList(data)$\\
$MinPts \leftarrow obtainMinPtsList(data, Eps)$\\
$counter \leftarrow 0 $\\
\For{$i\leq len(data)-1$}
{
$ClusterNum[i] \leftarrow$ DBSCAN$(Eps[i],MinPts[i])$\\
%$ClusterNum[i+1] \leftarrow$ DBSCAN$(Eps[i+1],MinPts[i+1])$\\
\If {$ClusterNum[i] == ClusterNum[i+1]$}
{
	$counter \leftarrow counter + 1 $\\ 
}
\Else
{
	$counter \leftarrow 0 $\\
}

	\If{$counter > 3$}
	{
		$n\leftarrow ClusterNum[i]$\\
		$left\leftarrow i$\\
		$right \leftarrow len(data)-1$\\
		\While{$start \leq end$}
		{
			$mid \leftarrow (start + end) / 2$\\
			\If{DBSCAN$(Eps[mid],MinPts[mid])<n$}
			{
				$right\leftarrow mid$\\
			}
			\ElseIf{DBSCAN$(Eps[mid],MinPts[mid])>n$}
			{
				$Left\leftarrow mid$\\
			}
			\Else
			{
			$\text{best index}\leftarrow mid$\\
			$k\leftarrow MinPts[\text{bext index}]$\\
			\textbf{return}  $k$
			}
		}
	}
}
\end{algorithm}

\subsection{To obtain Candidate Eps List}
To have a good performance on Multi-density datasets, multiple $Eps$ are required for Multi-density clustering. This part is to obtain the candidate $Eps$ list. In the step above, an adaptive $k$ is obtained based on the distribution property of the dataset $D$, which is utilized to determine the $k_{dis}$ value.

%\begin{figure}[htbp]
%  \centering
%  \includegraphics[width=\linewidth]{kdis}
%  \caption{$k_{dis}$ Curve}
%  \label{kdis}
%\end{figure}

To better locate candidate $Eps$, several algorithms (e.g., YADING\cite{dingYadingFastClustering2015}) have been proposed. These algorithms look for the inflection point (i.e., the part where the second-order differential is zero) of the $k_{dis}$ curve. In YADING\cite{dingYadingFastClustering2015} method, the flat part of the $k_{dis}$ curve indicates that the density of points is consistent, while the steep part indicates that the density of points is significantly different. However, our experimental results show that the candidate $Eps$ list derived from such methods does not give good clustering results. Furthermore, these methods introduce additional hyperparameters.

Inspired by YADING\cite{dingYadingFastClustering2015}, our AMD-DBSCAN proposes a new method of searching for candidate $Eps$ using the $k_{dis}$ frequency histogram, which is a frequency histogram of the $k_{dis}$ values. The peak of the $k_{dis}$ frequency histogram corresponds to the flat part of the $k_{dis}$ curve. Therefore, the $k_{dis}$ value of the peak of the histogram can be considered as a candidate $Eps$ because it ensures that there are a large number of points in this range that can be clustered into one class. To automatically distinguish between different peaks in the frequency histogram, the K-means algorithm is utilized to divide the $k_{dis}$ frequency histogram into different parts. And the clustering center of each part is considered as the candidate $Eps$. This process will allow more similar $k_{dis}$ values to be clustered in one class, and it is more representative of the $Eps$ of this part of the points.

The specific steps are as follows. The $k_{dis}$ value can be determined by using the adaptive $k$ derived from the last step and the $k_{dis}$ frequency histogram can be obtained by a series of $k_{dis}$ values. $N$ is considered as the number of clustering centers of the $k_{dis}$ values which is determined by observing the number of peaks of the $k_{dis}$ frequency histogram. And this $N$ is exactly equal to the number of clustering centers required for the K-means algorithm (i.e., K). Therefore, it is straightforward to utilize the K-means algorithm because K required by the K-means algorithm can be easily located. Moreover, our experiments in the latter section show that it is also effective because it has higher clustering accuracy. For example, as shown in Figure \ref{total}, it can be observed that there are three peaks in the frequency histogram. Therefore, $N$ is equal to 3, which means that there are three candidate $Eps$. The K-means algorithm with K equal to 3 can be used to cluster the $k_{dis}$ values. As shown in Figure \ref{total}, the $k_{dis}$ values corresponding to the center of the clustering result are considered as the candidate $Eps$. After the above process, $N$ candidate $Eps$ can be obtained, which means that most of the similar $k_{dis}$ values are clustered around these $N$ candidate $Eps$. These candidate $Eps$ can be used for Multi-density clustering.

The pseudo-code for obtaining candidate $Eps$ list is shown in Algorithm \ref{algorithm2}.
 \SetAlFnt{\small}
\begin{algorithm}
\SetKwData{Left}{left}\SetKwData{This}{this}\SetKwData{Up}{up}
\SetKwFunction{Union}{Union}\SetKwFunction{FindCompress}{FindCompress}
\SetKwInOut{Input}{input}\SetKwInOut{Output}{output}
\caption{To obtain Candidate Eps List}\label{algorithm2}
\Input{$data$\\
$k \leftarrow ParameterAdaptation(data)$}
\Output{$CandidateEpsList$: candidate $Eps$ list for Multi-density DBSCAN}
$distances \leftarrow sort(euclidean\_distances(data))$\\
$k_{dis} \leftarrow distances[k]$\\
$N \leftarrow$ Number of peak of the $k_{dis}$ frequency histogram\\
$CandidateEpsList \leftarrow$ K-means($k_{dis}$,K=$N$).centers\\
$\textbf{return}$  $CandidateEpsList$
\end{algorithm}

\subsection{Multi-density Clustering}
Once obtaining the candidate $Eps$ list, the next step is to perform Multi-density DBSCAN on the dataset $D$. Unlike the algorithm of YADING\cite{dingYadingFastClustering2015}, which sets all $MinPts$ to $k$ by default, our algorithm calculates $MinPts$ based on the distribution property of the dataset. $MinPts$ is calculated by the same algorithm as parameter adaptation, using the Equation \ref{minpts2}, which utilizes a candidate $Eps$ and $data$ to obtain $MinPts$. This process is encapsulated as the function $obtainMinPts(data, Eps)$. 

The specific process is as follows. The first step is to sort the obtained candidate $Eps$ list in ascending order, and then the obtainMinPts function is utilized to obtain the adaptive $MinPts$. Then, $data$, $Eps$, $MinPts$ are input into DBSCAN for clustering. After that, the $data$ that has been clustered is saved and is not clustered in the next loop. Repeat the above steps until all the $Eps$ are input into the model, and the remaining data is the noise. The whole clustering is finished.

The pseudo-code for Multi-density DBSCAN is shown in Algorithm \ref{algorithm3}.

 \SetAlFnt{\small}
\begin{algorithm}
\SetKwData{Left}{left}\SetKwData{This}{this}\SetKwData{Up}{up}
\SetKwFunction{Union}{Union}\SetKwFunction{FindCompress}{FindCompress}
\SetKwInOut{Input}{input}\SetKwInOut{Output}{output}
\caption{Multi-density DBSCAN}\label{algorithm3}
\Input{$data$\\
$EpsList \leftarrow obtainEpsCandidateList(data, k)$}
\Output{$cluster$ and $noise$}
\For{$Eps$ in $EpsList$}
{
	$MinPts \leftarrow obtainMinPts(data, Eps)$\\
	$cluster\leftarrow$ DBSCAN$(data, Eps, MinPts)$\\
	Mark $data$ with $labels$\\
	Remove $cluster$ from $data$\\
}
Mark $data$ as $noise$\\
$\textbf{return}$  $cluster$ and $noise$
\end{algorithm}

\subsection{Algorithm Complexity Analysis}
For a data set containing $n$ points, the complexity of DBSCAN is $O(n^2)$. By adopting the divide-and-conquer strategy, the complexity of DBSCAN is $O(n \log (n))$. For uniformity, the complexity of DBSCAN is denoted as $O(f(n))$. In the process of parameter adaptation, by using the binary search algorithm, the complexity is $O(\log (n))$, and thus the complexity of the whole parameter adaptation is $O(\log (n)f(n))$. In the process of locating the candidate $Eps$, the complexity of K-means algorithm is $O(n)$. The complexity of Multi-density DBSCAN in the final clustering process is also $O(f(n))$. In summary, the time complexity of AMD-DBSCAN is $O(\log (n)f(n)) +  O(n) + O(f(n)) =  O(\log (n)f(n))$.

\section{Experiment and Analysis}
In the experimental section, the datasets and evaluation metrics are introduced, then the AMD-DBSCAN algorithm is applied to Single-density and Multi-density datasets respectively to compare with different algorithms, and finally an ablation study is done to justify each step of our algorithm.

\subsection{Experiment platform and dataset}
The experimental platform in this paper is an AMD Ryzen 7 5800H processor with 16GB of RAM. To verify the performance of our algorithm, some classical algorithms are reproduced and compared with our algorithm in the same experimental platform.

In order to measure the difference in density between different clusters in a dataset, a new metric, the variance of the number of neighbors (VNN), is proposed. Equation \ref{eps} is utilized to obtain an $EpsList$. The first element of the EpsList is taken as $Eps1$, which is the average distance between all points and their nearest neighbors. And $Eps1$ is utilized as the search radius to search the number of neighbors within $Eps1$ of each data point. Then the variance of the number of neighbors (VNN) is shown in Equation \ref{VNN}.

\begin{equation}
\textbf{VNN} = Var(neighbors_{i}), 1 \leq i \leq n
\label{VNN}
\end{equation}

where $n$ represents the number of points of the dataset, and $neighbors_{i}$ represents the number of neighbors within $Eps1$ of $i^{th}$ point. 

The value of VNN is a metric that reflects the difference in density of the dataset. For Single-density datasets, the number of neighbors for each point within a given search radius is similar, resulting in a smaller value of VNN. In contrast, for Multi-density datasets with large density differences, the number of neighbors is large for dense clusters and small for sparse clusters, leading to a large value of VNN. Therefore, the larger the value of VNN is, the greater the density difference of datasets is. In this paper, datasets with VNN less than 10 are considered as Single-density datasets, while datasets with VNN greater than 10 are considered as Multi-density datasets, especially datasets with VNN greater than 100 are considered as extreme Multi-density datasets.

Several datasets are selected for our experiments from UCI\cite{frantiKmeansPropertiesSix2018}. And because the Multi-density datasets of extremely variable density are hard to obtain, some datasets are generated to better test the performance of our algorithm. The $make\_blobs1$ to $make\_blobs8$ datasets are generated by scikit-learn\cite{ScikitlearnMachineLearning}. $make\_blobs1$ serves as a base dataset where each cluster has the same amount of points and varies greatly in density. To test the robustness of the algorithm, some changes are made to the base dataset. 

The details of the datasets are shown in Table \ref{dataset}.

\begin{center}
\begin{table}[h]

\setlength{\tabcolsep}{3.5mm}{
  \caption{Dataset Properties}
  \label{dataset}
\begin{tabular}{ccccc}\toprule
\textbf{dataset} & \textbf{size} & \textbf{clusters} & \textbf{VNN}    & \textbf{Multi-density} \\ \midrule
Aggregation        & 788           & 7                 & 0.49                   & FALSE                                     \\
Compound           & 399           & 5                 & 5.66                    & FALSE                                      \\
D31                & 3100          & 31                & 1.76                     & FALSE                   \\
Flame              & 240           & 2                 & 1.12                      & FALSE                  \\
R15                & 600           & 15                & 1.62                     & FALSE                   \\
make\_blobs1       & 4998          & 6                 & 1726                     & TRUE                   \\
make\_blobs2       & 5048          & 6                 & 3743                    & TRUE                    \\
make\_blobs3       & 4998          & 6                 & 2953                    & TRUE                    \\
make\_blobs4       & 4998          & 6                 & 645                      & TRUE                  \\
make\_blobs5       & 4998          & 6                 & 4467                     & TRUE                   \\
make\_blobs6       & 4998          & 6                 & 127                      & TRUE                  \\
make\_blobs7       & 4998          & 6                 & 4100                    & TRUE                    \\
make\_blobs8       & 4998          & 6                 & 5600                     & TRUE                   \\
unbalance          & 6500          & 7                 & 17                     & TRUE                   \\
 \bottomrule                               
\end{tabular}}
\end{table}
\end{center}

\subsection{Evaluation Metrics}
There are two metrics utilized to evaluate the clustering results.
\begin{enumerate}
\item Normalized Mutual Information (NMI)\cite{danonComparingCommunityStructure2005} is utilized to evaluate the clustering results. NMI is between [0,1] and is used to measure the similarity of the clustering results, the larger the value, the better the clustering results.
\item The accuracy of clustering is utilized as the second metric. Since all datasets have labels, the accuracy is the percentage of correct labels after clustering.
\end{enumerate}

\subsection{Parameter Adaptation}
The experiment is designed to compare the execution time of our parameter adaptation approach with the traditional parameter adaptation approach PDDBSCAN\cite{luParallelAdaptiveDBSCAN2020} on different datasets. Two metrics are utilized, one is the shortest execution time and the other is the average execution time of 100 rounds of experiments. 

The detailed results of the experiments are illustrated in Table \ref{time}.
As can be seen from Table \ref{time}, after adopting the binary search algorithm, the speed of our algorithm is on average 4 times faster than PDDBSCAN\cite{luParallelAdaptiveDBSCAN2020}, which does not use the binary search algorithm. In particular, the speed of AMD-DBSCAN is much faster in some datasets where the same number of clusters occurs many times because the binary search algorithm can speed up the search process significantly. For example, in the $unbalance$ dataset, most of the points in this dataset are clustered in three classes. After using our algorithm, the execution time is reduced by nearly 93\%. Therefore, AMD-DBSCAN can speed up the execution time of the parameter adaptive process significantly on this kind of Multi-density dataset with a large number of points.

\begin{center}
\begin{table}[h]

\setlength{\tabcolsep}{3.5mm}{
  \caption{Comparisons of Execution Time}
  \label{time}
\begin{tabular}{ccccc}\toprule
\multicolumn{1}{l}{} & \multicolumn{2}{c}{\textbf{PDDBSCAN}} & \multicolumn{2}{c}{\textbf{AMD-DBSCAN}} \\\midrule
\textbf{dataset}   & \textbf{t\_min} & \textbf{t\_average} & \textbf{t\_min}  & \textbf{t\_average}  \\\midrule
Aggregation                      & 0.275            & 0.291     & 0.165           & 0.191               \\
Compound                         & 0.049            & 0.052     & 0.031           & 0.041              \\
D31                              & 2.550             & 2.820      & 1.350            & 2.330               \\
Flame                            & 0.042            & 0.045     & 0.021           & 0.024              \\
R15                              & 0.225            & 0.234     & 0.076           & 0.087              \\
unbalance                        & 95.658           & 97.782    & 6.407           & 6.552         \\
 \bottomrule     
\end{tabular}}
\end{table}
\end{center}

\begin{figure*}[htbp]
	\centering

	%\vspace{-1mm}
	    \subfigure{
		\rotatebox{90}{\scriptsize{\qquad\qquad\quad\textbf{\small ROCKA}}}
		\begin{minipage}[t]{1\linewidth}
			\centering
			\includegraphics[width=1\linewidth]{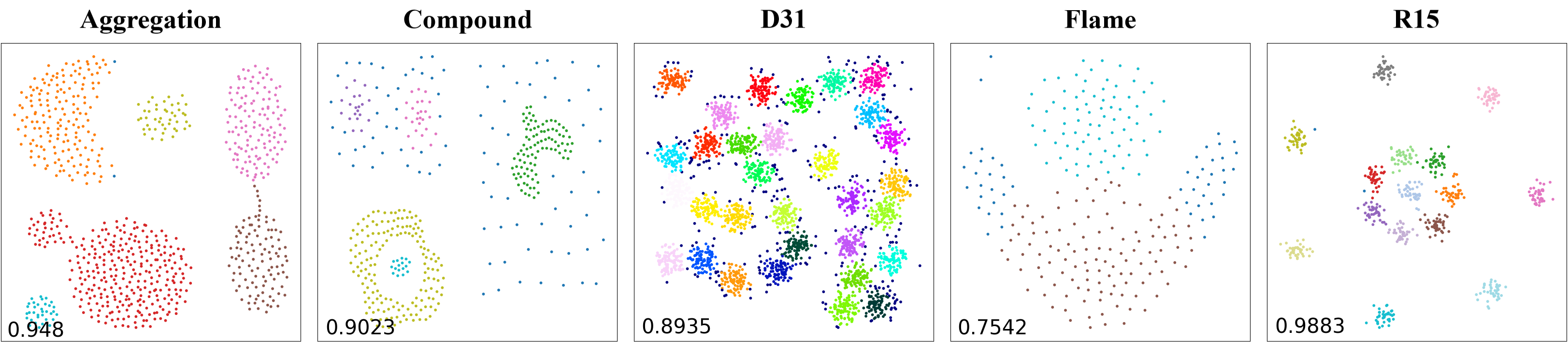}
		\end{minipage}
	}
	%\vspace{-1mm}
	    \subfigure{
		\rotatebox{90}{\scriptsize{\qquad\qquad\textbf{\small PDDBSCAN}}}
		\begin{minipage}[t]{1\linewidth}
			\centering
			\includegraphics[width=1\linewidth]{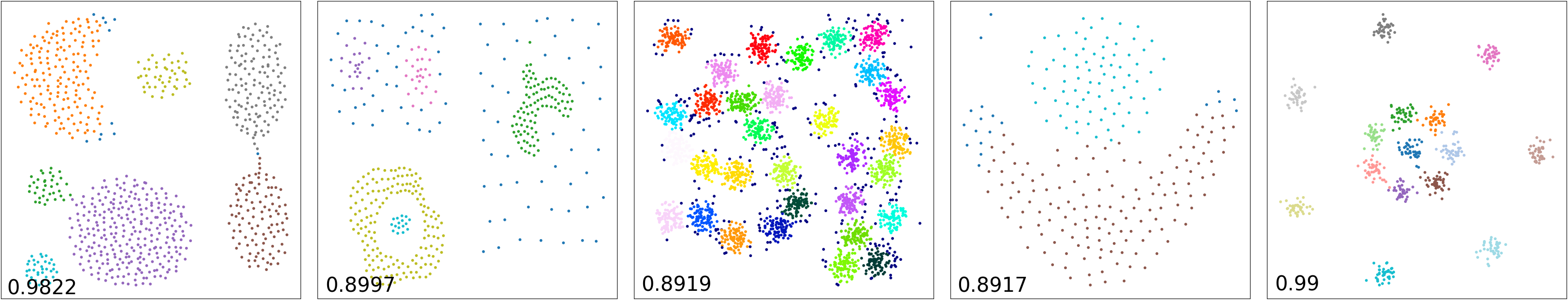}
		\end{minipage}
	}
    \subfigure{
		\rotatebox{90}{\scriptsize{\quad\qquad\textbf{\small AMD-DBSCAN}}}
		\begin{minipage}[t]{1\linewidth}
			\centering
			\includegraphics[width=1\linewidth]{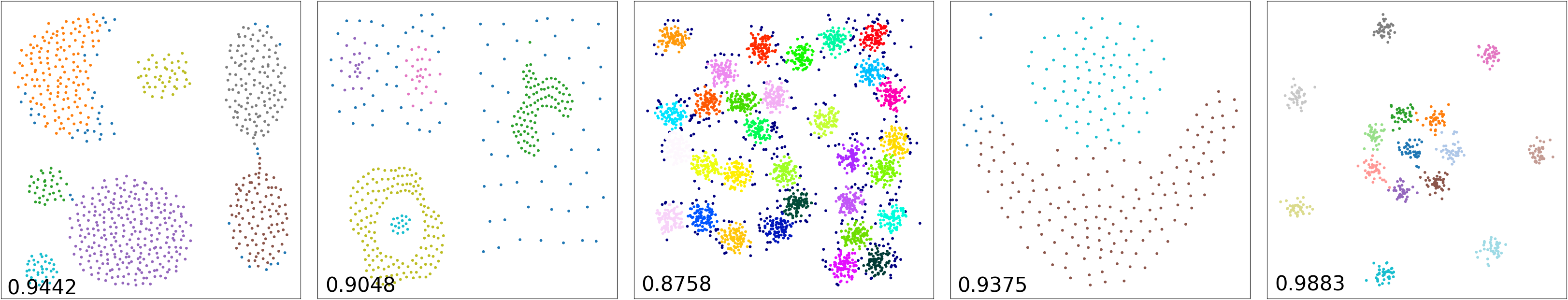}
		\end{minipage}
	}
	% 添加题注，即对这个图片的说明
	\caption{\textbf{Clustering results on Single-density datasets.} The number in the lower left corner represents the accuracy of clustering.}
	\label{fig1}
\end{figure*}

\subsection{Single-density Datasets}
\begin{figure}[htbp]
  \centering
  \includegraphics[width=\linewidth]{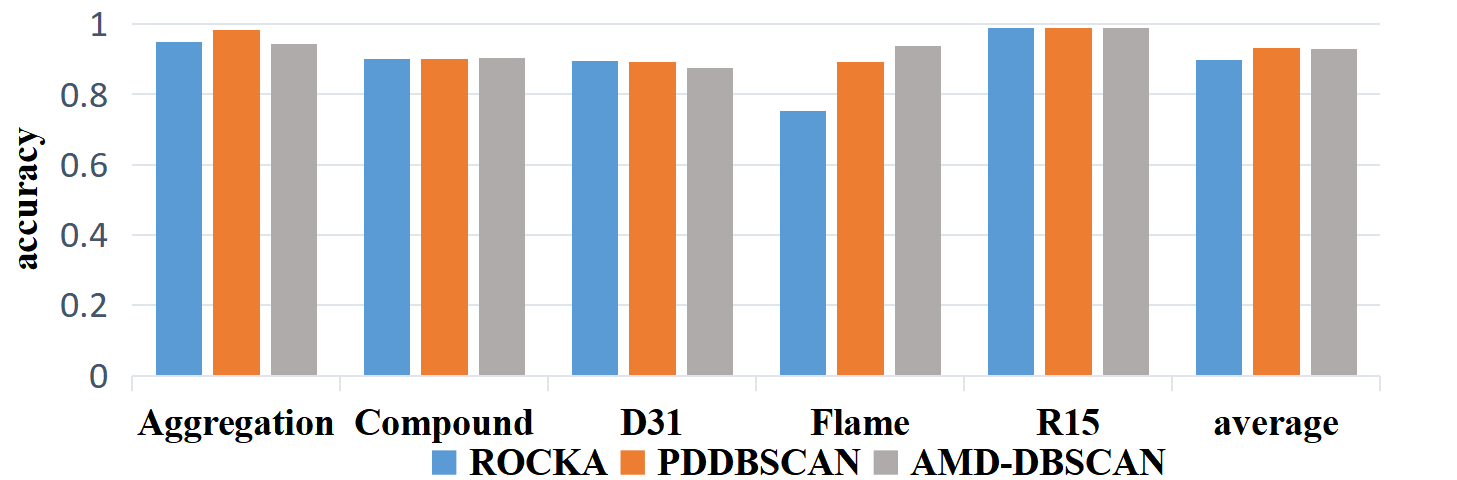}
  \caption{Comparisons of accuracy on Single-Density datasets}
  \label{cmp1}
\end{figure}

To compare the performamce on Single-density datasets (i.e., dataests with VNN less than 10) for our proposed AMD-DBSCAN, some Single-density clustering approaches (i.e., ROCKA\cite{liRobustRapidClustering2018} and PDDBSCAN\cite{luParallelAdaptiveDBSCAN2020}) are seleted. The experimental results are illustrated in Table \ref{result}, Figure \ref{cmp1} and the visualization effect is shown in Figure \ref{fig1}.

\begin{table}[htbp]
\setlength{\tabcolsep}{0.5mm}{
  \caption{Comparisons of Different Algorithm on Single-Density datasets}
 \label{result}
\begin{tabular}{c|ccc|ccc|ccc}\toprule
\multicolumn{1}{l}{} & \multicolumn{3}{c}{\textbf{ROCKA}}                                                                           & \multicolumn{3}{c}{\textbf{PDDBSCAN}}                                                                                       & \multicolumn{3}{c}{\textbf{AMD-DBSCAN}}                                                                         \\\midrule
\textbf{dataset}   & \textbf{accuracy}                      & \textbf{NMI}                           & \textbf{t(s)}        & \textbf{accuracy}                      & \textbf{NMI}                           & \textbf{t(s)}                 & \textbf{accuracy}                      & \textbf{NMI}                           & \textbf{t(s)}        \\\midrule
Aggregation         & 0.948          & 0.94          & \textbf{0.01} & \textbf{0.982} & \textbf{0.97} & 0.10    & 0.944          & 0.93          & 0.14\\

Compound              & 0.902          & \textbf{0.88} & \textbf{0.01} & 0.900          & 0.87          & 0.02  & \textbf{0.905} & 0.874         & 0.05 \\

D31                  & \textbf{0.894} & \textbf{0.88} & \textbf{0.14} & 0.892          & 0.88          & 1.13   & 0.876          & 0.87          & 1.24 \\

Flame                 & 0.754          & 0.57          & \textbf{0.01} & 0.892          & 0.66          & 0.02  & \textbf{0.938} & \textbf{0.75} & 0.04 \\

R15                  & 0.988          & 0.99          & \textbf{0.01} & \textbf{0.990} & \textbf{0.99} & 0.06  & 0.988          & 0.98          & 0.09 \\
 \bottomrule 
\end{tabular}} 
\end{table}

As shown in Figure \ref{cmp1}, compared to the other two algorithms, AMD-DBSCAN has an average accuracy of 1.6\% higher and achieves the best performance on the $Compound$ and $Flame$ datasets. Due to the process of parameter adaptation, our algorithm's execution time is longer than ROCKA\cite{liRobustRapidClustering2018} but still shorter than PDDBSCAN\cite{luParallelAdaptiveDBSCAN2020}.

Figure \ref{fig1} indicates that, for the $Aggregation$ dataset, AMD-DBSCAN considers more points as noise compared to the other algorithms but distinguishes all clusters. However, ROCKA\cite{liRobustRapidClustering2018} fails to distinguish two adjacent clusters.

As for the $Compound$ dataset, the $D31$ dataset, and the $R15$ dataset, the three algorithms have similar accuracy, which shows that they have good performance for datasets with lower VNN (i.e., datasets with uniform density distribution).

In the case of the $Flame$ dataset, which is comprised of only two clusters, the accuracy of AMD-DBSCAN is much higher than that of the other two algorithms, especially 18.3\% higher than that of ROKCA\cite{liRobustRapidClustering2018}. Because the points in this dataset are relatively scattered, many of them are considered as noise, which leads to low accuracy. 

Therefore, AMD-DBSCAN has a good performance on Single-density datasets.

\begin{figure*}[htbp]
	\centering

	%\vspace{-3mm}
	    \subfigure{
		\rotatebox{90}{\scriptsize{\quad\textbf{ YADING}}}
		\begin{minipage}[t]{1\linewidth}
			\centering
			\includegraphics[width=1\linewidth]{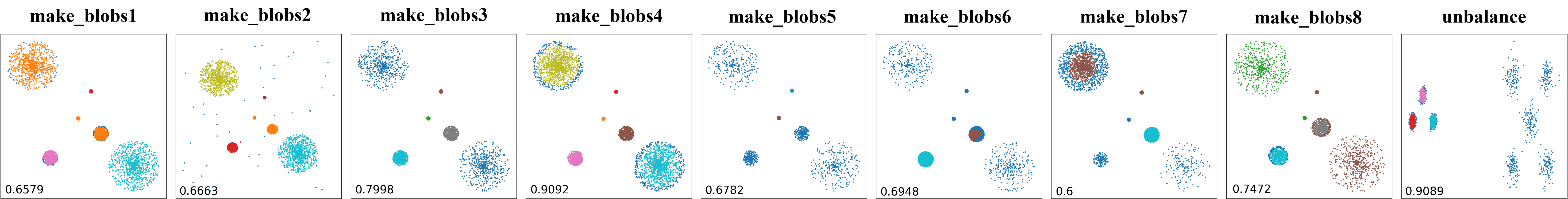}
		\end{minipage}
	}
	%\vspace{-3mm}
	    \subfigure{
		\rotatebox{90}{\scriptsize{\quad\textbf{AEDBSCAN}}}
		\begin{minipage}[t]{1\linewidth}
			\centering
			\includegraphics[width=1\linewidth]{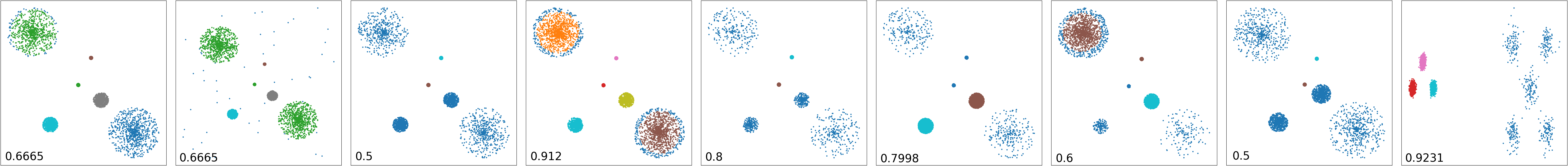}
		\end{minipage}
	}
    \subfigure{
		\rotatebox{90}{\scriptsize{\textbf{AMD-DBSCAN}}}
		\begin{minipage}[t]{1\linewidth}
			\centering
			\includegraphics[width=1\linewidth]{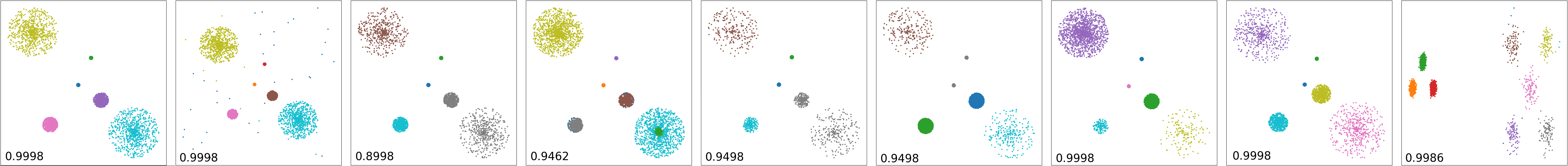}
		\end{minipage}
	}
	% 添加题注，即对这个图片的说明
	\caption{\textbf{Clustering results on Multi-density datasets.} The number in the lower left corner represents the accuracy of clustering.}
	\label{fig2}
\end{figure*}

\subsection{Multi-density Datasets}
\begin{figure}[htbp]
  \centering
  \includegraphics[width=\linewidth]{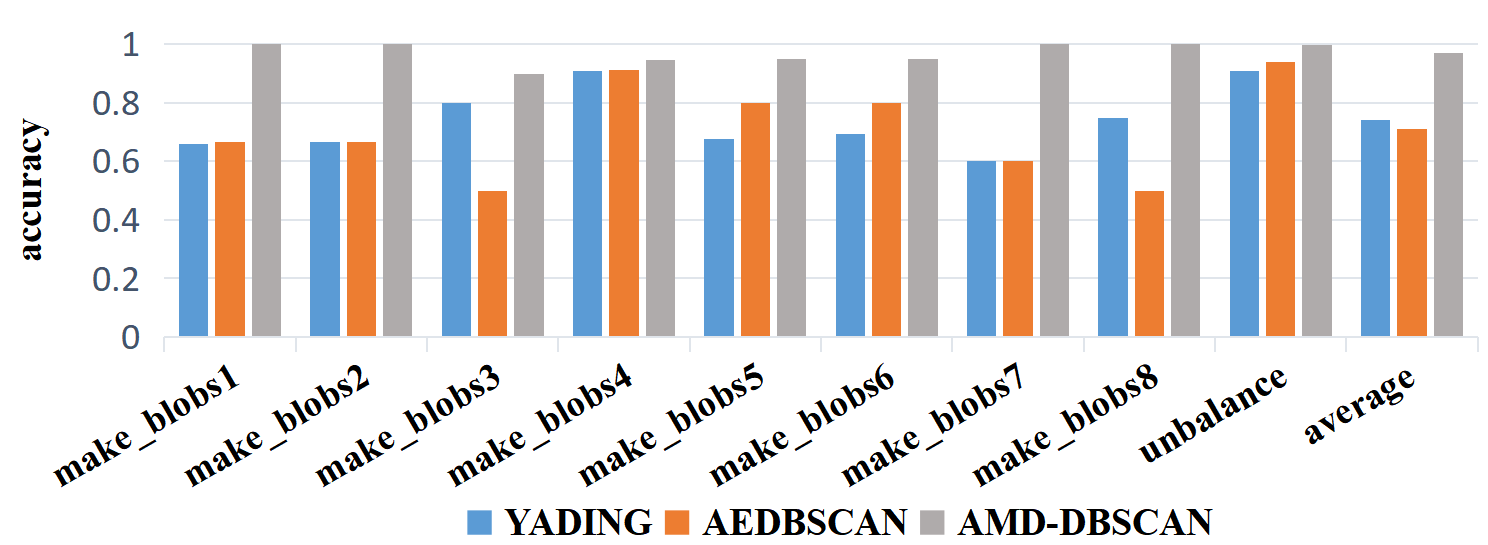}
  \caption{Comparisons of accuracy on Multi-density datasets}
  \label{cmp2}
\end{figure}

In order to evaluate the performamce on extreme Multi-density datasets (i.e., datasets with VNN larger than 100) for our proposed AMD-DBSCAN, some classical Multi-density clustering approaches (i.e., YADING\cite{dingYadingFastClustering2015}, AEDBSCAN\cite{mistryAEDBSCANAdaptiveEpsilon2021}) are selected for comparison experiments.
The experimental results are shown in Table \ref{resultmulti}, Figure \ref{cmp2} and the visualization effect is shown in Figure \ref{fig2}.

\begin{table}[htbp]
\setlength{\tabcolsep}{0.3mm}{
  \caption{Comparisons of Different Algorithm on Multi-density datasets}
 \label{resultmulti}
\begin{tabular}{c|ccc|ccc|ccc}\toprule
\multicolumn{1}{l}{} & \multicolumn{3}{c}{\textbf{YADING}}             & \multicolumn{3}{c}{\textbf{AEDBSCAN}}              & \multicolumn{3}{c}{\textbf{AMD-DBSCAN}}                                                                         \\\midrule
\textbf{dataset}   & \textbf{accuracy}  & \textbf{NMI}    & \textbf{t(s)} & \textbf{accuracy} & \textbf{NMI}   & \textbf{t(s)} & \textbf{accuracy}   & \textbf{NMI}  & \textbf{t(s)}  \\\midrule

make\_blobs1                           & 0.658             & 0.79          & \textbf{1.69}         & 0.666               & 0.90        & 3.69      & \textbf{1.000}     & \textbf{1.00}  & 3.62              \\

make\_blobs2                          & 0.666             & 0.84          & \textbf{1.33}         & 0.666               & 0.80         & 4.17        & \textbf{1.000}     & \textbf{0.99}  & 3.77           \\

make\_blobs3                           & 0.800             & \textbf{0.96} & \textbf{1.34}         & 0.500               & 0.75         & 4.28        & \textbf{0.900}     & 0.95           & 3.43           \\

make\_blobs4                          & 0.909             & 0.92         & \textbf{0.49}         & 0.912               & 0.91         & 4.14         & \textbf{0.946}     & \textbf{0.95}  & 3.06             \\

make\_blobs5                          & 0.678             & 0.64         & \textbf{2.52}         & 0.800               & 0.88       & 6.46         & \textbf{0.950}     & \textbf{0.93}  & 4.88                \\

make\_blobs6                          & 0.695             & 0.71          & \textbf{1.29}         & 0.800               & 0.88         & 5.22         & \textbf{0.950}     & \textbf{0.97}  & 12.52          \\

make\_blobs7                          & 0.600             & 0.69          & \textbf{1.66}         & 0.600               & 0.69         & 4.88            & \textbf{1.000}     & \textbf{1.00}  & 3.92        \\

make\_blobs8                           & 0.747             & 0.78          & \textbf{1.74}         & 0.500               & 0.75         & 4.602          & \textbf{1.000}     & \textbf{1.00}  & 3.64        \\

unbalance                             & 0.909             & 0.91          & 7.47                 & 0.938               & 0.95         & 12.11               & \textbf{0.999}     & \textbf{0.99}  & \textbf{7.23}\\ \bottomrule    
\end{tabular}} 
\end{table}

As illustrated in Figure \ref{cmp2}, AMD-DBSCAN has higher accuracy and NMI on Multi-density datasets of extremely variable density than the other two methods. Experimental results show that our AMD-DBSCAN has the best performance on the majority of datasets. Since having the parameter adaptation process, our algorithm's execution time is longer than YADING\cite{dingYadingFastClustering2015} but still shorter than AEDBSCAN\cite{mistryAEDBSCANAdaptiveEpsilon2021}.

As shown in Figure \ref{fig2}, different colors are represented in different clusters. For the $make\_blobs1$ dataset, clusters with small radius have high density while clusters with large radius have low density. AMD-DBSCAN achieves the best clustering performance, while the other two algorithms consider the points at the edge of each cluster as noise. AEDBSCAN\cite{mistryAEDBSCANAdaptiveEpsilon2021} even considers a sparse cluster as noise, resulting in low accuracy. 

As for the $make\_blobs2$ dataset, it is proven that our AMD-DBSCAN is highly resistant to noise by the evaluation of adding noise to the base dataset.

In terms of the $make\_blobs3$ and $make\_blobs4$ datasets, the number of points of different clusters is modified. AMD-DBSCAN has the highest accuracy. However, the other two algorithms are not able to distinguish those sparse clusters and the points at the edge of the clusters.

In the case of from the $make\_blobs5$ to $make\_blobs8$ datasets, the distribution between each cluster is changed and the whole dataset is more sparsely distributed. For these sparse points, the other two algorithms consider them as noise, while AMD-DBSCAN can distinguish these points from the denser ones and cluster them correctly.

The $unbalance$ dataset is a Multi-density dataset in which the vast majority of points are in the three high-density clusters on the left, while the five on the right are low-density clusters. Since the number of these low-density clusters is small, these clusters are considered as noise. The experimental results show that the other two algorithms can only distinguish three high-density clusters. In contrast, AMD-DBSCAN can completely distinguish eight clusters with high accuracy. 

To sum up, AMD-DBSCAN improves accuracy by an average of 24.7\% over the other two algorithms on Multi-density datasets of extremely variable density, which proves that the candidate $Eps$ obtained by our algorithm and their corresponding $MinPts$ are adapted to the distribution of the dataset.

\begin{table*}[htbp]
\setlength{\tabcolsep}{3mm}{
  \caption{Ablation study}
 \label{compresult}
\begin{tabular}{c|ccc|ccc|ccc|cc}\toprule
\multicolumn{1}{l}{} & \multicolumn{3}{c}{\textbf{AMD-DBSCAN}}         & \multicolumn{3}{c}{\textbf{Test1(k=4)}}    & \multicolumn{3}{c}{\textbf{Test2(k=n/2)}}    & \multicolumn{2}{c}{\textbf{Test3(no K-means)}}               \\\midrule
\textbf{dataset} & \textbf{accuracy} & \textbf{NMI}   & \textbf{clusters} & \textbf{accuracy} & \textbf{NMI}   & \textbf{clusters} & \textbf{accuracy} & \textbf{NMI} & \textbf{clusters} & \textbf{accuracy} & \textbf{NMI}   \\\midrule
Aggregation        & 0.944             & 0.931          & 7                                               & 0.675             & 0.728          & 24                                              & 0.346                                     & 0.000                                & 1                                               & \textbf{0.957}    & \textbf{0.939}                                              \\
Compound           & \textbf{0.905}    & \textbf{0.874} & 5                                               & 0.782             & 0.831          & 5                                               & 0.396                                     & 0.082                                & 1                                               & 0.789             & 0.867                                                       \\
D31                & \textbf{0.876}    & \textbf{0.870} & 31                                              & 0.555             & 0.671          & 101                                             & 0.032                                     & 0.000                                & 1                                               & 0.813             & 0.820                                                        \\
Flame              & \textbf{0.938}    & 0.748          & 2                                               & 0.925             & \textbf{0.808} & 2                                               & 0.638                                     & 0.000                                & 1                                               & 0.925             & 0.718                                                       \\
R15                & \textbf{0.988}    & \textbf{0.985} & 15                                              & 0.620             & 0.660          & 26                                              & 0.105                                     & 0.104                                & 2                                               & 0.983             & 0.980                                                        \\
make\_blobs1       & \textbf{1.000}    & \textbf{1.000} & 6                                               & 0.875             & 0.878          & 35                                              & 0.325                                     & 0.411                                & 3                                               & 0.825             & 0.807                                                       \\
make\_blobs2       & \textbf{1.000}    & \textbf{0.992} & 6                                               & 0.904             & 0.919          & 10                                              & 0.165                                     & 0.114                                & 3                                               & 0.851             & 0.834                                                      \\
make\_blobs3       & 0.900             & 0.949          & 5                                               & 0.834             & 0.871          & 24                                              & 0.580                                     & 0.690                                & 4                                               & \textbf{0.966}    & \textbf{0.961}                                               \\
make\_blobs4       & \textbf{0.946}    & \textbf{0.955} & 7                                               & 0.618             & 0.789          & 51                                              & 0.250                                     & 0.000                                & 1                                               & 0.834             & 0.916                                                        \\
make\_blobs5       & \textbf{0.950}    & \textbf{0.973} & 5                                               & 0.603             & 0.777          & 17                                              & 0.547                                     & 0.592                                & 4                                               & 0.899             & 0.888                                                        \\
make\_blobs6       & \textbf{0.950}    & \textbf{0.973} & 5                                               & 0.718             & 0.772          & 28                                              & 0.450                                     & 0.259                                & 2                                               & 0.678             & 0.830                                                      \\
make\_blobs7       & \textbf{1.000}    & \textbf{1.000} & 6                                               & 0.907             & 0.902          & 19                                              & 0.630                                     & 0.688                                & 4                                               & 0.756             & 0.705                                                   \\
make\_blobs8       & \textbf{1.000}    & \textbf{1.000} & 6                                               & 0.840             & 0.898          & 23                                              & 0.400                                     & 0.503                                & 2                                               & 0.965             & 0.958                                                    \\
unbalance          & \textbf{0.999}    & \textbf{0.998} & 9                                               & 0.629             & 0.685          & 92                                              & 0.323                                     & 0.322                                & 2                                               & 0.997             & 0.992           \\                                           
\bottomrule       
\end{tabular}} 
\end{table*}

\subsection{Ablation Study}
In order to verify the rationality of each step of AMD-DBSCAN, three sets of ablation study are designed. 

First, Test1 verifies that the adaptive $k$ derived from parameter adaptation is valid. In YADING\cite{dingYadingFastClustering2015} and ROCKA\cite{liRobustRapidClustering2018}, $k$ is taken as 4 by default. In Test1, $k$ is taken to be 4, and then continue to complete our clustering process. The experimental results shown in Table \ref{compresult} indicate that $k$ obtain by the parameter adaptation is instructive for determining $k_{dis}$ value because it can locate an adaptive $k$ according to the distribution characteristics of the dataset. However, if $k$ is constant by default 4, the accuracy of DBSCAN decreases by an average of 20.8\%.

Test2 together with Test1 proves the theory presented in section 3.2, that is, a large $k$ leads to a small number of clusters, while a small $k$ results in a large number of clusters. In Test2, $n$ represents the number of points in the dataset, and $k$ is taken to be $n/2$, which is a very large value compared to that in Test1. The comparisons of the number of clusters from Test1 and Test2 show that when $k$ is small, the number of clusters is large, and when $k$ is large, the number of clusters is small, which proves that an appropriate $k$ is important because it affects the clustering results significantly.

Test3 is to verify that the candidate $Eps$ obtained by the K-means algorithm is valid. The K-means algorithm in AMD-DBSCAN is replaced by YADING's algorithm\cite{dingYadingFastClustering2015}, and then continue to complete our clustering process. Test3 proves that the algorithm proposed in this study for processing the $k_{dis}$ frequency histogram using the K-means algorithm is valid. This is because K can be easily determined by directly observing the $k_{dis}$ frequency histogram. And K-means is effective at clustering one-dimensional data like $k_{dis}$ values\cite{wagstaffConstrainedKmeansClustering2001}. Experimental results show that the clustering accuracy of our algorithm is on average 8.3\% better compared to YADING's algorithm \cite{dingYadingFastClustering2015} without using the K-means algorithm.

In summary, the ablation study shows that each step of AMD-DBSCAN is necessary because it guides the selection of better parameter pairs thus improving clustering effect.

\section{Discussion}
The current AMD-DBSCAN implementation is single-threaded. From the discussion of the AMD-DBSCAN algorithm, it is clear that the AMD-DBSCAN implementation can be easily parallelized because it can perform adaptive operations in multiple partitions at the same time. In specific, some partitioning strategies and distributed structures (e.g., MapReduce\cite{mohammedApplicationsMapReduceProgramming2014}) can be utilized in our model. Hence, our model can be applied to cluster in each small partition and finally merge the clustering results. In this way, our model can divide the large-scale dataset into different small partitions, use a parameter adaptation process to determine the parameter pairs on each small partition, and use parallelization to speed up this process. Therefore, our model can be applied to large-scale datasets. As a result, the performance of AMD-DBSCAN can be significantly improved by parallelization.

\section{Conclusion}
In this paper, an adaptive Multi-density DBSCAN algorithm (AMD-DBSCAN) with high robustness and efficiency is proposed. First, an improved parameter adaptation method is proposed to locate $k$. The binary search algorithm is utilized to speed up the adaptive process and the experimental results show that the speed of our algorithm is on average 4 times faster than the traditional methods. Second, instead of calculating the inflection points of the $k_{dis}$ curves, multiple candidate $Eps$ can be obtained by using the $k_{dis}$ frequency histogram and the K-means algorithm. And compared to other approaches, AMD-DBSCAN requires only one hyperparameter that is easily determined. In addition, AMD-DBSCAN improves the algorithm of Multi-density clustering so that there is an adaptive parameter pair for each density cluster. Furthermore, the variance of the number of neighbors (VNN) is proposed to measure the difference in density. The experimental results show that the clustering accuracy of AMD-DBSCAN is 24.7\% higher than other algorithms on average on Multi-density datasets of extremely variable density and is 1.6\% higher on average on Single-density datasets.

%\clearpage

%\bibliographystyle{ACM-Reference-Format}
\bibliographystyle{IEEEtran}

\bibliography{DBSCAN}

\end{document}